\documentclass[twocolumn,twocolappendix,lowtilde]{aastex63}

\usepackage[T1]{fontenc}        % font encoding
\usepackage{amsmath,mathtools}  % mathematical notation
\usepackage{hyperref}           % links

\hypersetup{colorlinks=true,linkcolor=black,citecolor=black,filecolor=black,urlcolor=black}

\DeclarePairedDelimiter{\paren}{\lparen}{\rparen}
\DeclarePairedDelimiter{\ave}{\langle}{\rangle}
\DeclarePairedDelimiter{\abs}{\lvert}{\rvert}
\newcommand{\parenpow}[3]{\paren[#1]{#2}^{\!\!#3\,}}
\newcommand{\ee}{\mathrm{e}}
\newcommand{\ii}{\mathrm{i}}
\newcommand{\dd}{\mathrm{d}}

\newcommand{\kb}{k_\mathrm{B}}

\newcommand{\rad}{\mathrm{rad}}
\newcommand{\seconds}{\mathrm{s}}
\newcommand{\ms}{\mathrm{ms}}
\newcommand{\yr}{\mathrm{yr}}
\newcommand{\cm}{\mathrm{cm}}
\newcommand{\km}{\mathrm{km}}
\newcommand{\g}{\mathrm{g}}
\newcommand{\msun}{M_\odot}
\newcommand{\mev}{\mathrm{MeV}}
\newcommand{\gauss}{\mathrm{gauss}}

\newcommand{\ye}{Y_\mathrm{e}}
\newcommand{\ro}{\mathrm{Ro}}
\newcommand{\rol}{\mathrm{Ro}_\ell}
\newcommand{\vconv}{v_\mathrm{conv}}
\newcommand{\vturb}{v_\mathrm{turb}}
\newcommand{\vave}{v_\mathrm{ave}}
\newcommand{\vavevec}{\vec{v}_\mathrm{ave}}
\newcommand{\bdip}{B_\mathrm{dip}}
\newcommand{\fohm}{f_\mathrm{ohm}}
\newcommand{\feff}{f_\mathrm{eff}}
\newcommand{\fout}{F_\mathrm{out}}
\newcommand{\fconv}{F_\mathrm{conv}}
\newcommand{\lconv}{L_\mathrm{conv}}
\newcommand{\lnu}{L_\mathrm{nu}}
\newcommand{\fnu}{F_\nu}
\newcommand{\fnue}{F_{\nu_e}}
\newcommand{\fnup}{F_{\bar{\nu}_e}}
\newcommand{\fnuh}{F_{\nu_x}}
\newcommand{\vprime}{\vec{v}^{\,\prime}}
\newcommand{\bprime}{\vec{b}^{\,\prime}}
\newcommand{\tauc}{\tau_\mathrm{c}}
\newcommand{\kcrit}{k_\mathrm{crit}}
\newcommand{\kfast}{k_\mathrm{fast}}
\newcommand{\ofast}{\omega_\mathrm{fast}}
\newcommand{\lmax}{\ell_\mathrm{max}}
\newcommand{\lturb}{L_\mathrm{turb}}
\newcommand{\re}{\mathrm{Re}}
\newcommand{\rem}{\mathrm{Re}_\mathrm{m}}
\newcommand{\pr}{\mathrm{Pr}}
\newcommand{\prm}{\mathrm{Pr}_\mathrm{m}}

\newcommand{\code}[1]{\texttt{#1}}

\shorttitle{Pulsar and Magnetar Magnetic Fields}
\shortauthors{White, Burrows, Coleman, Vartanyan}

\begin{document}

% Title
\title{On the Origin of Pulsar and Magnetar Magnetic Fields}

% Author information
\author{Christopher~J.~White}
\affiliation{Department of Astrophysical Sciences, Princeton University, Princeton, NJ, 08544, USA}
\author[0000-0002-3099-5024]{Adam~Burrows}
\affiliation{Department of Astrophysical Sciences, Princeton University, Princeton, NJ, 08544, USA}
\author{Matthew~S.~B.~Coleman}
\affiliation{Department of Astrophysical Sciences, Princeton University, Princeton, NJ, 08544, USA}
\author[0000-0003-1938-9282]{David~Vartanyan}
\affiliation{Department of Astronomy, University of California, Berkeley, CA, 94720, USA}

% Abstract
\begin{abstract}

  In order to address the generation of neutron star magnetic fields, with particular focus on the dichotomy between magnetars and radio pulsars, we consider the properties of dynamos as inferred from other astrophysical systems. With sufficiently low (modified) Rossby number, convective dynamos are known to produce dipole-dominated fields whose strength scales with convective flux, and we argue that these expectations should apply to the convective proto-neutron stars at the centers of core-collapse supernovae. We analyze a suite of three-dimensional simulations of core collapse, featuring a realistic equation of state and full neutrino transport, in this context. All our progenitor models, ranging from $9\ \msun$ to $25\ \msun$, including one with initial rotation, have sufficiently vigorous proto-neutron-star convection to generate dipole fields of order ${\sim}10^{15}\ \gauss$, if the modified Rossby number resides in the critical range. Thus, the magnetar/radio pulsar dichotomy may arise naturally in part from the distribution of core rotation rates in massive stars.

\strut

\end{abstract}

% Introduction
\section{Introduction}
\label{sec:introduction}

Observations have revealed a dichotomy in the population of neutron stars (NSs):\ flaring, slowly rotating magnetars with strong inferred magnetic fields are distinct from faster-rotating radio pulsars with weaker fields. Explaining these two populations remains an open problem in the modeling of core-collapse supernovae and in subsequent NS evolution. While a large number of core-collapse simulations successfully match observed supernova properties and leave behind bound NS remnants \citep{Couch2014,Lentz2015,Vartanyan2019,Muller2019,Burrows2019,Burrows2020,Vartanyan2020,Bollig2021}, it is less clear how such modeling can account for the diversity of NS magnetic fields seen in Nature.

To date only approximately thirty magnetars have been discovered \citep{Olausen2014},\footnote{See \url{http://www.physics.mcgill.ca/~pulsar/magnetar/main.html} for the updated catalog.} though their transient nature implies that $10\%$ or more of all young NSs could fall into this population \citep{Kaspi2017}. The cataloged spin periods range from $2$ to $12$\ seconds (s), and their surface dipolar fields as calculated from the periods and period derivatives \citep[see][for a recent review]{Igoshev2021} are $10^{13}\text{--}10^{15}\ \gauss$. Magnetars are young, with most having characteristic spin-down ages of less than $10^4\ \yr$. They are distributed closer to the galactic plane than radio pulsars \citep{Kaspi2017}, and eight are associated with known supernova remnants \citep{Olausen2014}.

This population stands in contrast to that of the isolated radio pulsars, of which there are over $3000$ cataloged members \citep{Manchester2005}.\footnote{See \url{http://www.atnf.csiro.au/research/pulsar/psrcat} for the updated catalog.} The bulk of pulsar periods are between $0.1$ and a few seconds, and young pulsars have periods of $0.5\ \seconds$ or less \citep{Narayan1987,Emmering1989}. Observations are well matched by initial periods being $300 \pm 150\ \ms$ \citep{FaucherGiguere2006}. Pulsar surface dipole magnetic fields span $10^{11}\text{--}10^{13}\ \gauss$, with the youngest objects falling in the $10^{12}\text{--}10^{13}\ \gauss$ range \citep{Kaspi2016}.

The key to this dichotomy lies in the magnetic fields with which NSs are endowed during core collapse. While magnetars also have notably slower rotation rates, this is consistent with being born rotating as fast as (or possibly faster than) their less magnetized counterparts and subsequently rapidly losing angular momentum due to strong magnetic braking. Indeed, population modeling by \citet{Jawor2022} shows that magnetar initial periods must be less than $2\ \seconds$.

One idea is that the field strengths seen in newly born NSs are simply the result of amplification due to flux freezing during collapse \citep{Ferrario2006,Ferrario2008,Hu2009}, with the diversity of outcomes reflecting the range of main-sequence stellar fields. \Citet{Spruit2008} counters that the change in radius from that of the stellar core to that of the NS could not amplify reasonable fields to magnetar levels. On the other hand, \citet{Cantiello2016} note asteroseismic modeling indicates some massive stars may well be sufficiently magnetized to become magnetars through this simple mechanism. Still, flux freezing can be only a part of the explanation, since, as \citet{Spruit2008} notes, a strong pre-existing field would couple the core to the envelope, reducing core angular momentum below what is seen in young pulsars.

Alternatively, field amplification can proceed more dynamically. Rotation is likely an important ingredient, with the interplay of rotation and magnetic fields being an important part of pre-collapse stellar evolution \citep{Heger2005} as well as appearing in the dynamics of the proto-neutron star (PNS) itself \citep{Nagakura2020}. The growth of field via compression, linear and exponential dynamos, and the magnetorotational instability \citep[MRI,][]{Velikhov1959,Balbus1991} might produce jet-launching magnetars in the cores of long-duration gamma-ray bursts \citep{Obergaulinger2020,Aloy2021,Obergaulinger2021a,Obergaulinger2021b} and/or hypernovae \citep{Burrows2007}. These models generally rely on rapid rotation, and indeed with such short rotational timescales even very small seed fields can be amplified to magnetar levels in less than $1\ \seconds$ \citep{Mosta2015}.

A related mechanism for amplification is that of a convective dynamo. Given the expectation of convective regions in PNSs shortly after bounce \citep{Burrows1987,Burrows1988}, \citet{Duncan1992} estimate that strong convective dynamo action during this time can greatly amplify magnetic fields, explaining magnetars in particular.

Within the convective dynamo framework, there happens to be a natural explanation for a dichotomy of outcomes. As shown by \citet{Christensen2006} and \citet{Olson2006} in the context of local planetary systems, there are simple scaling laws obeyed by these systems. In particular, the existence of a dipole field depends critically on the Rossby number relating the rotational and convective timescales. Dynamos that are sufficiently rapidly rotating yield a saturated field dominated by a dipole whose strength scales with the power provided by buoyancy, while slower rotation leads to a broader multipolarity.

Fortuitously for our purposes, these results are generally found to hold so long as the magnetic Reynolds number is sufficiently large, as it is in the PNS case, and they are otherwise largely independent of dimensionless numbers associated with diffusive effects, including the magnetic Reynolds, Ekman, Prandtl, and magnetic Prandtl numbers \citep{Christensen2006}. While momentum, thermal, and magnetic diffusivities can vary wildly across astrophysical dynamos, PNS Rossby numbers are similar to those of Solar-System bodies. The relationship between convective flux and dipolar field strength has been successfully extended from planetary bodies to low-mass, rapidly rotating stars, including both T~Tauri stars and old M~dwarfs \citep{Christensen2009}. We seek in this paper to extend this connection further to encompass PNS magnetic field generation.

Some recent work has focused on simulating field growth in the PNS context, with varying strategies for simplifying the problem. \Citet{Franceschetti2020} see dynamo action with a prescribed, stationary velocity field and a polytropic PNS. \Citet{ReboulSalze2021} explore MRI effects with an incompressible (and, therefore, non-convective) model. With an imposed heat flux and the anelastic equations, \citet{Raynaud2020} find strong field growth in rapidly rotating models. \Citet{Masada2021} use a nuclear equation of state and drive convection by fixing the lepton fraction profile, whose gradient drives PNS convection in Nature; they see large-scale eddies more akin to meridional flows than standard small-scale turbulence, which go to the center of the model and generate a dipolar field even with slow rotation. None of these models directly include neutrino transport, and the physics they include and conclusions they reach are varied.

\begin{figure}
  \centering
  \includegraphics[width=3.35in]{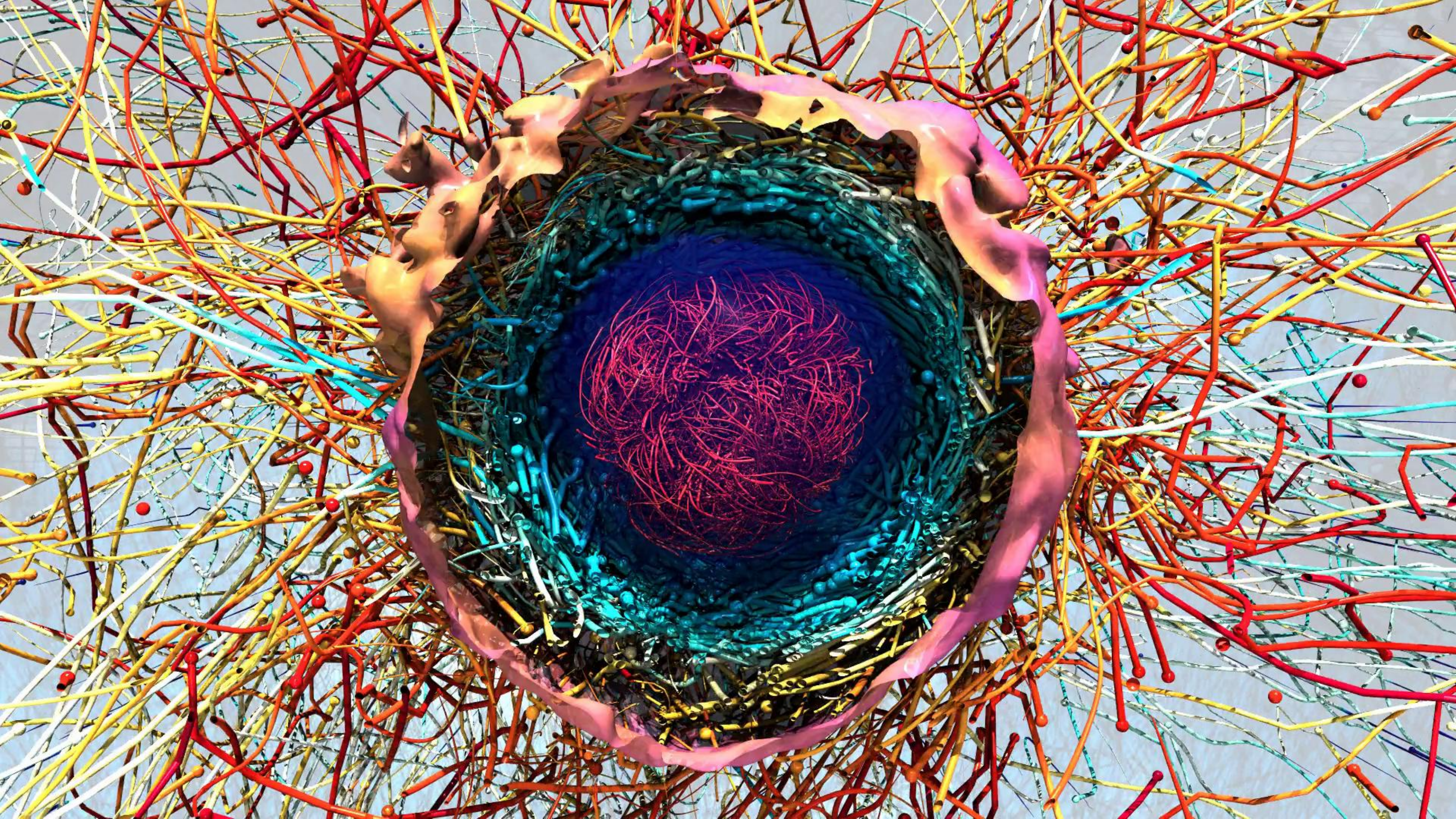}
  \caption{Three-dimensional rendering of the PNS from a simulation of a $25\ \msun$ performed by \citet{Burrows2020}, including a $10^{9.5}\ \g\ \cm^{-3}$ density isosurface, as well as tracer particles following the inner PNS convection and the outer turbulent motions. The colors distinguish various values of electron fraction $\ye$. \label{fig:3d_pns}}
\end{figure}

We believe it is worthwhile to explore PNS dynamos from the heretofore neglected angle of Rossby number and convective flux scaling, applying the lessons learned from large suites of dynamo simulations to the PNS regime. In this light, we analyze a number of core-collapse simulations performed using the \code{Fornax} code \citep{Skinner2019,Vartanyan2019}. These include three-dimensional models with progenitors ranging in mass from $9\ \msun$ to $25\ \msun$ \citep{Burrows2020}, where one new $9\ \msun$ model (unpublished) features initial rotation. While these simulations do not include magnetic fields, their advanced treatment of the other relevant processes of core collapse, including neutrino transport, neutrino-matter coupling, and the equation of state, allow us to use them to confidently anchor our analysis to physically realistic and self-consistent conditions. All of our models attain vigorous convection in and around a PNS; Figure~\ref{fig:3d_pns} shows a rendering of one snapshot with velocity tracer particles and a density isosurface delineating the PNS.

In \S\ref{sec:dynamo_scaling}, we analyze our models in terms of convective scaling relations. Section~\ref{sec:convection} explores the convective region in more detail. We discuss the implications of this analysis in \S\ref{sec:discussion}. Details about diffusivities and dimensionless numbers can be found in Appendix~\ref{sec:diffusivities}.

% Convective Dynamo Scaling
\section{Convective Dynamo Scaling}
\label{sec:dynamo_scaling}

\begin{figure}
  \centering
  \includegraphics{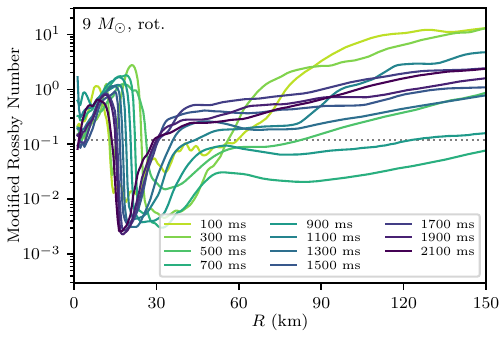}
  \caption{Profiles of modified Rossby number $\rol$, altered to account for the scale of turbulence as in \citet{Christensen2006} and \citet{Olson2006}, at various times in the rotating $9\ \msun$ model. This model has a slow initial spin period in its center of ${\sim}60\ \seconds$ and a final average spin period of ${\sim}20\ \ms$. Only with $\rol \lesssim 0.12$ is a strong dipole expected. In this model the low Rossby number region around $R = 30\ \km$ lies near, but mostly outside, the region of PNS convective luminosity. The region from $50\ \km$ to $150\ \km$, despite having a low absolute rotation rate (Figure~\ref{fig:angular_velocity}), has a Rossby number below the critical value for approximately $500\ \ms$. \label{fig:rossby}}
\end{figure}

The Rossby number $\ro = \vconv / \Omega R$ characterizes the rate of rotation relative to eddy turnover times, with faster rotation corresponding to lower Rossby numbers. According to \citet{Christensen2006} and \citet{Olson2006}, convective dynamos with a modified Rossby number
\begin{equation}
  \rol = \frac{\bar{\ell}}{\pi} \cdot \frac{\vconv}{\Omega R} \lesssim 0.12
\end{equation}
should produce strong dipolar fields. Here, we take $\vconv$ to be the radial contribution to $\vturb$,
\begin{equation}
  \vconv = \ave[\big]{(v^r - \vave^r)^2}_\Omega^{1/2},
\end{equation}
thus more unambiguously associating it with convection. The average angular scale $\bar{\ell}$ is defined in Appendix~\ref{sec:diffusivities}. The evolution of radial profiles of $\rol$ is shown in Figure~\ref{fig:rossby}. The $9\ \msun$ rotating model is always below the critical threshold near $30\ \km$, and for several hundred milliseconds is below the threshold exterior to this region.

In the $\rol \lesssim 0.12$ regime, \citet{Christensen2009} provide a scaling relation for the strength of the saturated dipole field, expressed in our notation as
\begin{equation} \label{eq:dipole}
  \frac{\bdip^2}{8 \pi} = c \fohm \feff \parenpow{\big}{\ave{\rho} \fout^2}{\,1/3}.
\end{equation}
Here $c = 0.63$ is a constant of proportionality fit by \citet{Christensen2009}; we assume $\fohm = 1$ (all dissipation is ultimately ohmic). $\fout$ is the total flux (convective and neutrino) at the outer edge of the convection zone. Given the density and temperature scale heights $H_\rho = -\rho / (\dd \rho / \dd r)$ and $H_T = -T / (\dd T / \dd r)$, as well as the convective flux $\fconv$, the efficiency factor is
\begin{equation} \label{eq:feff}
  \feff = \ave[\Bigg]{\parenpow{\bigg}{\frac{H_\rho}{H_T}}{2/3} \parenpow{\bigg}{\frac{\rho}{\ave{\rho}}}{1/3} \parenpow{\bigg}{\frac{\fconv}{\fout}}{2/3}}.
\end{equation}
All averages are volume averages over the convective region.

\begin{figure*}
  \centering
  \includegraphics{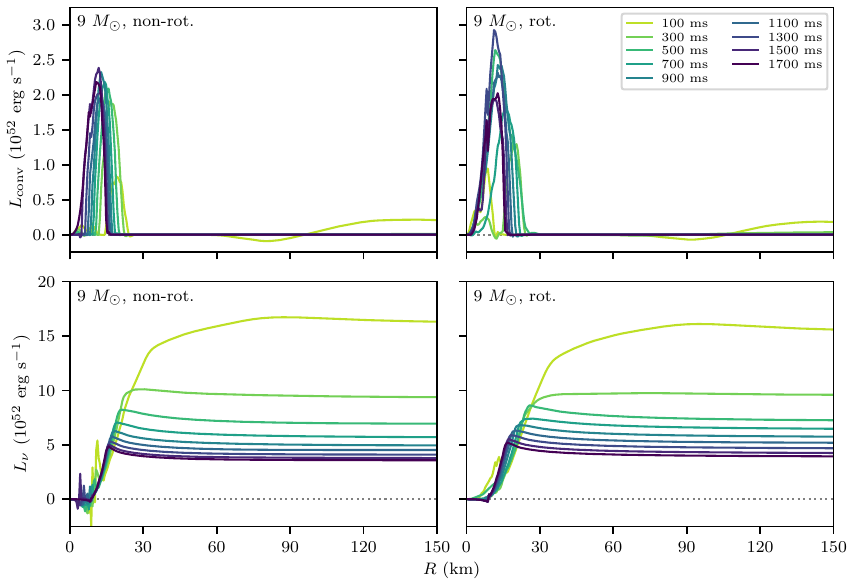}
  \caption{Lagrangian luminosities for the non-rotating (left) and rotating (right) $9\ \msun$ models at different post-bounce times. The top panels show the convective luminosities, which can drive a turbulent dynamo in the PNS. The bottom panels show the contributions of neutrinos, summed over all species, which dominate the total luminosity (note the different vertical scales). \label{fig:luminosities}}
\end{figure*}

In order to apply this scaling, we need to measure the fluxes in our simulations. Because they are not static, we are careful to do this in the reference frame in which the material at a given radius and time has a vanishing average radial momentum. The convective enthalpy flux comes from terms involving the kinetic energy, internal energy, and pressure:
\begin{equation} \label{eq:fconv}
  \fconv = \ave[\bigg]{\paren[\bigg]{\frac{1}{2} \rho \vturb^2 + u + p} \vconv^r}_\Omega.
\end{equation}
For comparison, we also consider the neutrino flux, which is simply the angle-averaged radial component in this frame:
\begin{equation}
  \fnu = \ave{\fnue^r + \fnup^r + \fnuh^r}_\Omega,
\end{equation}
where the three neutrino species followed are the $\nu_e$; $\bar{\nu}_e$; and the sum of the $\nu_\mu$, $\bar{\nu}_\mu$, $\nu_\tau$, and $\bar{\nu}_\tau$ neutrinos aggregated into one species. Multiplying by $4 \pi R^2$ yields the corresponding luminosities $\lconv$ and $\lnu$, which we compare for the $9\ \msun$ models in Figure~\ref{fig:luminosities}. At all times, the convective flux is strongly peaked inside the PNS convective region, as characterized further in \S\ref{sec:convection}. The modestly rotating $9\ \msun$ model has convection going all the way to the center. The neutrino luminosities grow to be much larger, and they are relatively constant in radius outside $30\ \km$.

\begin{figure*}
  \centering
  \includegraphics{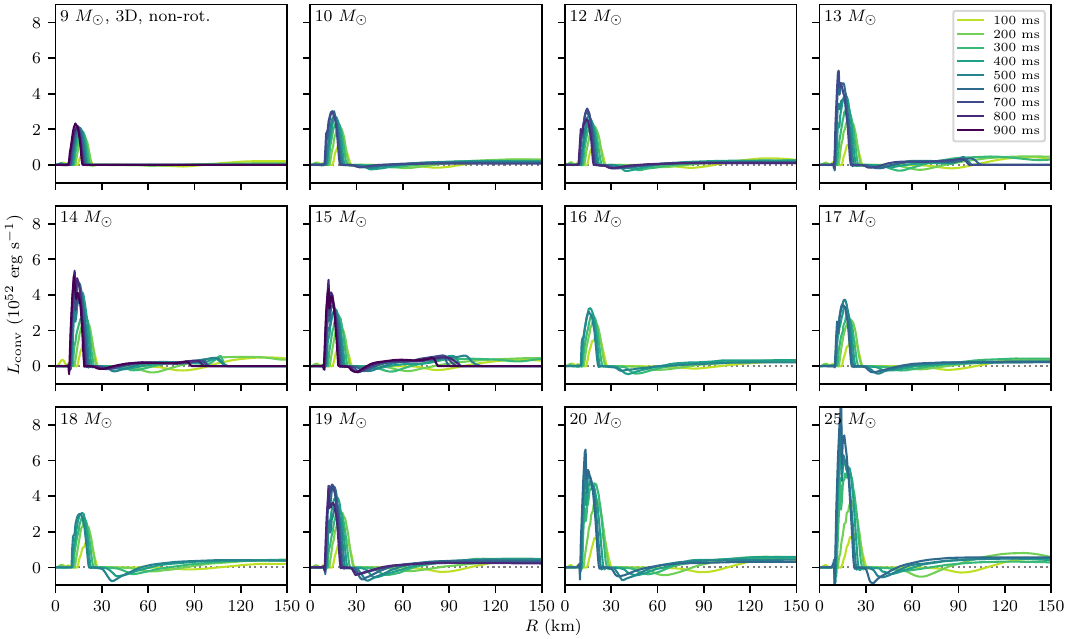}
  \caption{Profiles of Lagrangian convective luminosities at different post-bounce times. There is a general trend of more vigorous convection with larger progenitor masses. Furthermore, the strength of convection generally increases over time for the durations of these simulations, going to approximately one second after bounce in some cases. \label{fig:convection_3d}}
\end{figure*}

In Figure~\ref{fig:convection_3d}, in order to provide a sense of how strongly convection scales with progenitor mass, we plot the convective luminosities for all other 3D models from \citet{Burrows2020} and the unpublished rotating $9\ \msun$ model. The peak luminosity grows with progenitor mass, reaching values approximately $4$ times higher in the $25\ \msun$ case relative to the $9\ \msun$ case.

\begin{figure*}
  \centering
  \includegraphics{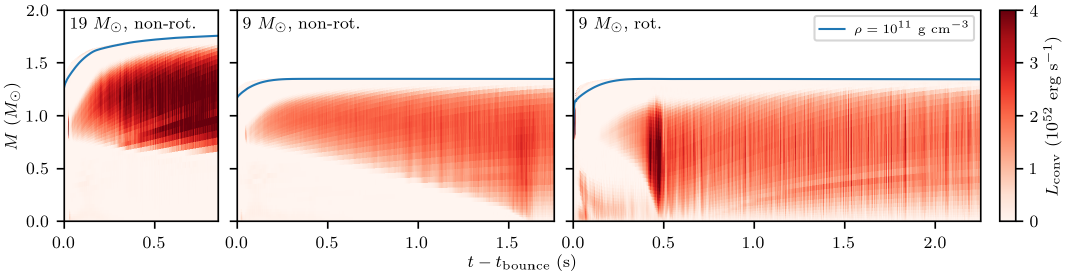}
  \caption{Mass-time diagrams of convective luminosity for three illustrative models. At high (left) and low (center) progenitor masses, as well as intermediate masses (not shown), the inner half of the mass does not become convective within one second of bounce (but is likely to do so at later times, and does so at ${\sim}1.7\ \seconds$ for the non-rotating $9\ \msun$ model). However, these deep layers are convective much earlier in the $9\ \msun$ rotating model (right), with something akin to meridional circulation seen in its deep interior. The solid lines denote the boundary of the PNS. In all our models, there is a ${\sim}0.1\ \msun$ radiative layer on the PNS ``surface.'' \label{fig:convection_mass_time}}
\end{figure*}

Several features of PNS convective flux are more easily illustrated in mass coordinates. Figure~\ref{fig:convection_mass_time} shows a mass-time diagram of $\lconv$ for three models. For the $9\ \msun$ non-rotating model, there is appreciable convection in the center of the PNS only after ${\sim}1.7\ \seconds$. For the other non-rotating models, it generally takes as much time or longer to achieve this state. In contrast, the moderately rotating $9\ \msun$ model has convective flux throughout most of the mass as early as ${\sim}0.5\ \seconds$.\footnote{The behavior in the deep interior vis-\`a-vis PNS convection of this unpublished $9\ \msun$ (rotating) model is novel, and it may be associated with only a small subset of progenitors. It is the first of its kind (other groups' spherical 3D simulations model the deep interior in 1D or 2D) and needs to be checked. However, its rotational profiles are likely robust and it is these we employ here. In addition, its novel feature---the early onset of full-core convection---is only quantitatively different from the behavior of other 3D models, in that they too eventually achieve full-core convection, only later. For instance, the non-rotating $9\ \msun$ model is fully convective by $1.7\ \seconds$ after bounce. Such global convection of the core and possible core meridional convection \citep{Masada2021} will increase the scale height and the convective timescale, thereby decreasing the Rossby number in the way we highlight in this paper as potentially important for magnetic field generation in the modestly rotating case.} Furthermore, we consistently find a radiative region in the outer ${\sim}0.1\ \msun$ of the PNS. This is independent of progenitor mass, PNS mass, and rotation. The solid lines in Figure~\ref{fig:convection_mass_time} denote the PNS boundary, defined arbitrarily as the $10^{11}\ \g\ \cm^{-3}$ surface. This ${\sim}0.1\ \msun$ layer is important in that any field generated in the convective region must be transported through it in order to contribute to the more readily observable NS surface and exterior fields.

\begin{figure}
  \centering
  \includegraphics{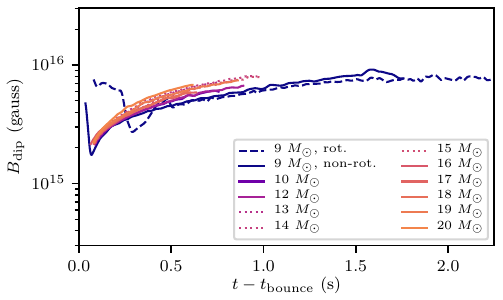}
  \caption{Predicted saturated dipole strengths of varying progenitors, as given by the low-Rossby-number scaling with convective flux found in \citet{Christensen2009}. For each snapshot in time, the value plotted is the predicted dipole field assuming conditions in that snapshot persist until saturation. The environment is such that one expects dipole field strengths of over $10^{15}\ \gauss$, given sufficiently low Rossby number and sufficient time to reach saturation. For clarity, the lines have been smoothed with a Gaussian filter of standard deviation $10\ \ms$. \label{fig:expected_dipole}}
\end{figure}

Defining the convective region to end where the convective luminosity drops below $1\%$ of its peak value, we can apply Equations~\eqref{eq:dipole}, \eqref{eq:feff}, and~\eqref{eq:fconv} to our models to find the predicted saturated dipole field assuming the \citeauthor{Christensen2009}\ scaling relations hold. This can be done for each snapshot, and the results as functions of time are shown in Figure~\ref{fig:expected_dipole}. As there is snapshot-to-snapshot noise in the lines, for clarity we smooth them with a Gaussian filter ($10\ \ms$ standard deviation).

For all our models, the scaling predicts dipole field strengths of at least $5 \times 10^{15}\ \gauss$, with slightly larger values at higher progenitor masses. That is, there is enough convective flux to drive the creation of magnetar-level fields. Note that while assuming the scaling relations from \citet{Christensen2009} hold in this regime of $\rol \lesssim 0.12$, they only give the saturated field strength. That is, Figure~\ref{fig:expected_dipole} shows the time evolution of the saturated value, not the growth of the field itself.

Given how long the saturated strength remains large (beyond the time limits of any of our simulations), it seems there is a good chance for saturation to be reached. For illustrative numbers, we consider the rotating $9\ \msun$ model $500\ \ms$ after bounce. In this snapshot, there are $1030$ sound crossing times (center to $\rho = 10^{11}\ \g\ \cm^{-3}$ surface) per second; the fastest azimuthal velocities correspond to $136$ rotations per second (and thus there are $136\ \rol$ convective turnovers per second). Comparing these numbers to the $14$ $\ee$-foldings needed to grow the field by a factor of $10^{6}$, we infer that $1\text{--}2\ \seconds$ of sustained dynamo action could well be sufficient to generate magnetar fields from weak progenitor fields. We return to the issue of saturation while employing a linear analysis in \S\ref{sec:convection}.

It is important to note that we are not predicting such strong fields are ubiquitous. The scaling relation only predicts field strengths in the limit that the modified Rossby number is sufficiently low. For slower rotation, weaker, multipolar fields are predicted, with dynamos near the critical value possibly showing polarity reversals akin to those found in Earth's magnetic field. Our single rotating model has regions of sufficiently low Rossby number (Figure~\ref{fig:rossby}). Much of the convection, including the peak, is found only interior to $20\ \km$ (Figure~\ref{fig:luminosities}, upper right panel).

\begin{figure}
  \centering
  \includegraphics{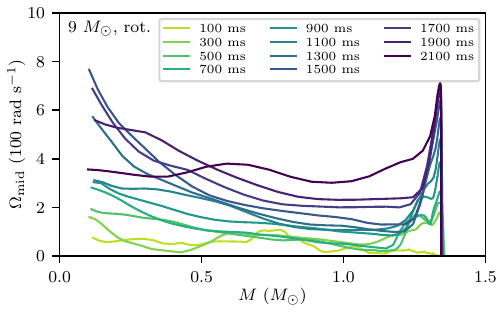}
  \caption{Midplane angular velocity profiles at various times in the rotating $9\ \msun$ model. The rapidly rotating material piles in the outer region of the PNS. Rotation rates for this model are relatively modest ($\lesssim 400\ \rad\ \seconds^{-1}$ in the PNS, and $\lesssim 150\ \rad\ \seconds^{-1}$ immediately exterior to this). \label{fig:angular_velocity}}
\end{figure}

As we are particularly interested in the broader implications of rotation as it applies to dynamos, we take a more detailed look at the rotation profile of the rotating $9\ \msun$ model. Figure~\ref{fig:angular_velocity} shows the radial profile of the angular velocity in the midplane at various times, plotted against mass coordinate. Note that the ordinate indicates the angular velocity at each abscissa, not the average interior to each abscissa. Angular velocity shows a strong peak near $1.3\ \msun$ (${\sim}20\text{--}40\ \km$) that grows with time. This is the effect of material that has accreted later, conserving its angular momentum and, thus, increasing its angular velocity, while not being able to efficiently penetrate the outermost envelope of the PNS to spin up the core at this early phase. Since the more coarse gridding in the deep interior can lead to large uncertainties near the coordinate singularity, we exclude the inner $0.1\ \msun$ of material from the plot.

\begin{figure}
  \centering
  \includegraphics{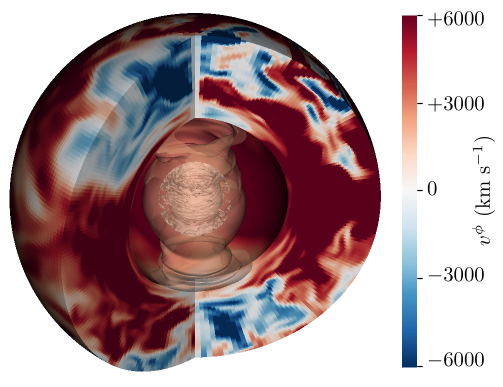}
  \caption{Azimuthal velocity of the unpublished rotating $9\ \msun$ model $300\ \ms$ after bounce. The outer sphere is located at $R = 100\ \km$. The inner $50\ \km$ has been removed and replaced by the $1000\ \km\ \seconds^{-1}$ and $2500\ \km\ \seconds^{-1}$ isosurfaces. \label{fig:3d_vph}}
\end{figure}

This rotation is not uniformly distributed in latitude. Figure~\ref{fig:3d_vph} renders the three-dimensional azimuthal velocity field inside $R = 100\ \km$, $300\ \ms$ after bounce. The inner $50\ \km$ have been cut away and replaced with the $1000\ \km\ \seconds^{-1}$ and $2500\ \km\ \seconds^{-1}$ isosurfaces.

% Proto-Neutron-Star Convection
\section{Proto-Neutron-Star Convection}
\label{sec:convection}

In order for field amplification to occur, there must be a source of free energy in the system, which can then drive appropriate motion in the fluid. In the inner PNS context, there is indeed turbulence across a large range of scales, driven by convection. Here convection is not driven by an entropy gradient; indeed, entropy increases outward in the cores of most of our models. Instead, it is driven by the lepton composition gradient, which is maintained by the process of deleptonization as neutrinos escape from the outside inward \citep{Keil1996,Dessart2006,Nagakura2020}. We note that \citet{Olson2006} find that models with heat generated within the convective zone (as would happen with radioactive decay in a planetary mantle) do not display the same low-Rossby-number asymptotics as those with heat sourced only at boundaries. As we do not have explicit heat sources, nor are we considering the gain region outside the PNS core, we believe the latter models from \citet{Olson2006} are more applicable. Still, lepton-gradient-driven convection is different enough that further study is warranted.

\begin{figure*}
  \centering
  \includegraphics{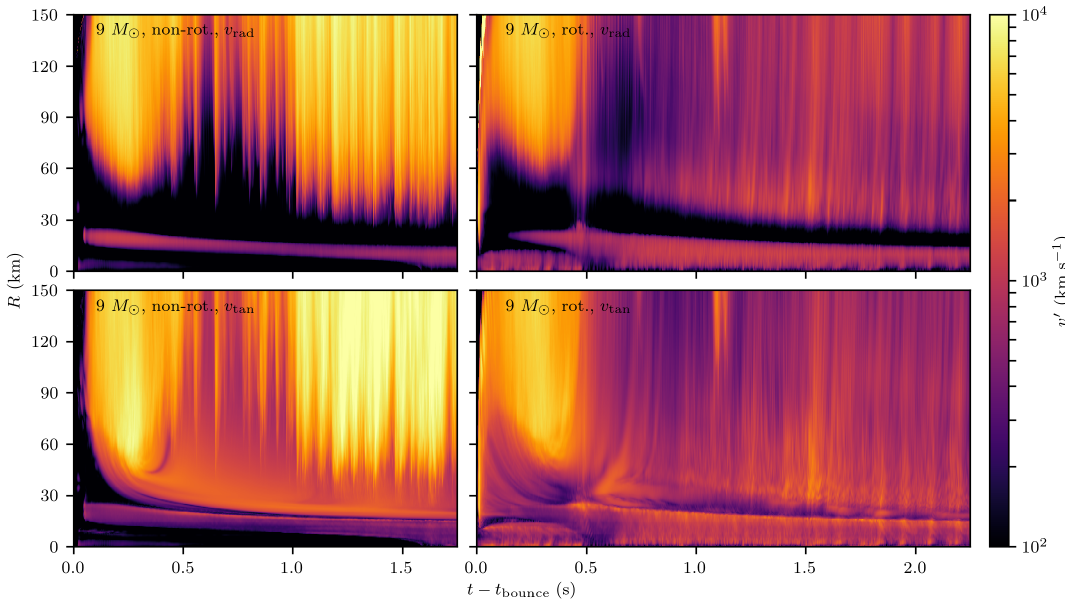}
  \caption{RMS radial (top) and tangential (bottom) turbulent velocities in the non-rotating (left) and rotating (right) $9\ \msun$ models. The quantities are averaged in angle after subtracting bulk radial motion, and, in the case of the rotating model, bulk azimuthal velocity. Large turbulent velocities trace convective regions. The rotating model differs qualitatively from its non-rotating counterpart in that its convection extends all the way to the center of the grid much earlier than other models. Whether the behavior seen for this rotating model survives and is robust (even if for only a subset of progenitors) is the subject of future work. \label{fig:velocity}}
\end{figure*}

Figure~\ref{fig:velocity} highlights turbulent regions with the radial and tangential contributions to $\vturb$ in the two $9\ \msun$ models, non-rotating and rotating. The non-rotating case shows the same qualitative features as found with the other progenitors:\ a well-defined convective region at the outer layers of the PNS (initially around $20\text{--}30\ \km$), which slowly moves inward as the core cools and contracts, and another region of turbulent velocities exterior to this, going out beyond $150\ \km$. The latter region is not convective at its base (interior to $60\ \km$), as can be seen by the lack of turbulent radial velocities. This is consistent with the lack of convective flux here, as shown in Figure~\ref{fig:luminosities}. Still, the material in this region, which has passed through the shock, maintains large velocities for extended periods of time, and thus may be relevant for magnetic field generation.

The rotating case displays several qualitative differences. The PNS convection zone can still be seen in radial velocity, though by around $500\ \ms$ after bounce its inner boundary extends all the way to the center. Furthermore, there is a distinct zone of turbulent motion in the inner $10\ \km$ of the PNS at early times. In general there are no parts of the region we are examining (interior to $150\ \km$, and interior to the shock, which can be seen as the boundary between zero and nonzero velocities in the upper left of the lower right panel in Figure~\ref{fig:velocity}) that do not have significant turbulent velocities. We emphasize that bulk radial and azimuthal motion has been subtracted, leaving to show only smaller-scale motions.

\begin{figure}
  \centering
  \includegraphics{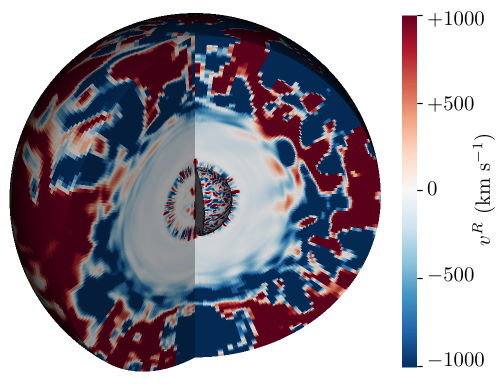}
  \caption{Three-dimensional rendering of the radial velocity distribution of the rotating $9\ \msun$ model $300\ \ms$ after bounce. The outer sphere is located at $R = 100\ \km$. The inner wedge has a radius of $20\ \km$, coincident with the PNS convection zone at this time. \label{fig:3d_vr}}
\end{figure}

A three-dimensional visualization of radial velocity in a single snapshot is provided in Figure~\ref{fig:3d_vr}. This shows the rotating $9\ \msun$ model $300\ \ms$ after bounce, out to $R = 100\ \km$. The large velocities from $50\ \km$ to $100\ \km$ are visible, and one can clearly see the convective zone near $R = 20\ \km$. In this latter region, the convective cells' heights are a substantial fraction of the thickness of the turbulent layer itself.

\begin{figure}
  \centering
  \includegraphics{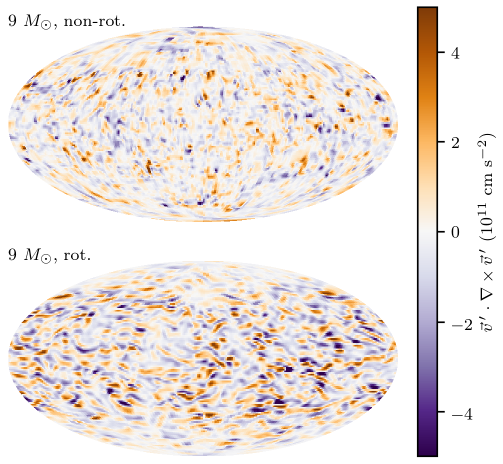}
  \caption{Velocity (kinetic) helicity $\vprime \cdot \nabla \times \vprime$ in the non-rotating (top) and rotating (bottom) $9\ \msun$ models, on a Mollweide projection of the $R = 20\ \km$ slice, $300\ \ms$ after bounce. The bulk radial motion has been subtracted. The helicity field of the rotating model is qualitatively different, with more extreme values and elongated structures near the midplane. \label{fig:helicity_spheres}}
\end{figure}

\begin{figure}
  \centering
  \includegraphics{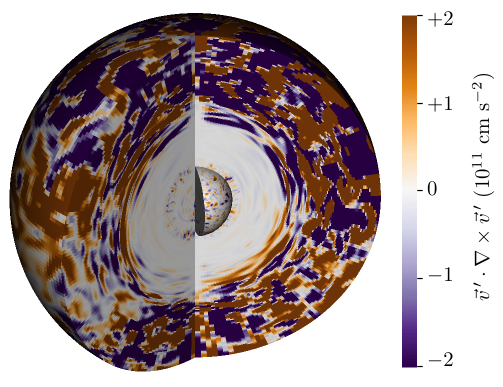}
  \caption{Cutaway 3D view of velocity (kinetic) helicity $\vprime \cdot \nabla \times \vprime$ in the rotating $9\ \msun$ model, $300\ \ms$ after bounce. The outer sphere is located at $R = 100\ \km$, while the inner wedge has a radius of $20\ \km$, coincident with the PNS convection zone at this time. \label{fig:3d_helicity}}
\end{figure}

An important aspect of dynamos is the (kinetic) helicity in the small-scale velocity field $\vprime$, given by
\begin{equation}
  h = \vprime \cdot \nabla \times \vprime.
\end{equation}
Figure~\ref{fig:helicity_spheres} shows the helicity on spherical surfaces ($R = 20\ \km$) $300\ \ms$ after bounce in the two $9\ \msun$ models, where the rotating model has azimuthally elongated features near the midplane. The three-dimensional rendering in Figure~\ref{fig:3d_helicity} shows the same quantity for $R < 100\ \km$, with a wedge at $R = 20\ \km$, for the rotating model. The non-convective outer layers of the PNS, with essentially no helicity, separate the turbulent regions with large helicities.

\begin{figure*}
  \centering
  \includegraphics{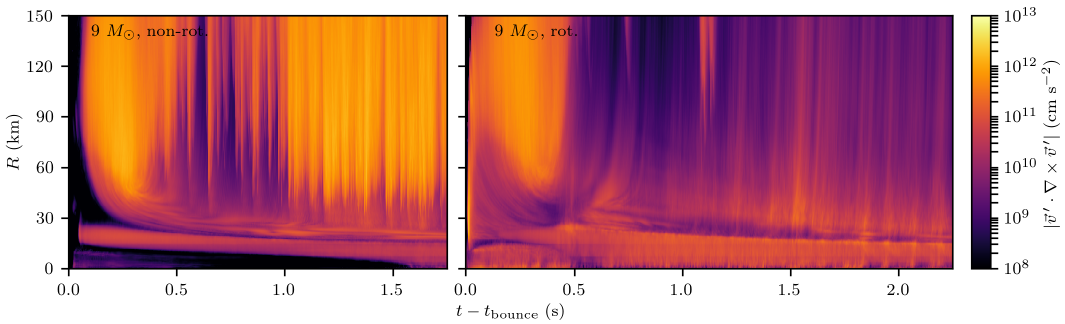}
  \caption{Velocity (kinetic) helicity $\vprime \cdot \nabla \times \vprime$ in the non-rotating (left) and rotating (right) $9\ \msun$ models, averaged in angle. The bulk radial motion has been subtracted. For the rotating model, the average azimuthal velocity at each radius and latitude has been subtracted as well. \label{fig:helicity}}
\end{figure*}

In general, field growth should be aided by larger magnitudes of helicity. This can be seen by writing the induction equation for the mean magnetic field as
\begin{equation}
  \frac{\partial\ave{\vec{B}}}{\partial t} = \nabla \times \paren[\big]{\ave{\vec{v}} \times \ave{\vec{B}} + \alpha \ave{\vec{B}} - (\eta + \beta) \nabla \times \ave{\vec{B}}}
\end{equation}
\citep{Charbonneau2014}, where $\eta$ is the magnetic diffusivity, $\alpha = -\tauc \ave{h} / 3$ drives mean field amplification on small scales, $\beta = \tauc \ave{(\vprime)^2} / 3$ incorporates the fact that small-scale turbulent reconnection acts as an effective diffusivity, and $\tauc$ is a correlation timescale for the turbulence.\footnote{More generally, $\alpha = -\tauc / 3 \times (\ave{h} - \rho^{-1} \bprime \cdot \nabla \times \bprime)$, where $\bprime$ is the subgrid RMS magnetic field in the mean-field decomposition.} As shown in Figure~\ref{fig:helicity}, regions of large helicity broadly trace those of large tangential turbulent velocity (see Figure~\ref{fig:velocity}).

Rotating models already have extra kinetic energy from their bulk motions, compared to equal-progenitor-mass non-rotating models. Further, they can supply free energy sourced from their differential rotation. The above discussion shows that, even beyond these points, rotating models experience more turbulent and helical motion, which makes them more favorable for dynamo action.

The amenability for dynamo growth can be seen more quantitatively in a linear mode analysis. As analyzed by \citet{Shibata2021} in an NS context (see \citet{Brandenburg2005} for a general treatment), a mode with frequency $\omega$ (positive imaginary parts unstable), wavenumber $k$, and wavenumber $k_\parallel$ parallel to the rotation axis has the dispersion relation
\begin{equation}
  \ii \, \omega = \eta k^2 \pm \sqrt{\alpha^2 k^2 - \ii \alpha k_\parallel S_\Omega},
\end{equation}
where $S_\Omega = r (\dd\Omega / \dd r)$ measures the shear in cylindrical radius $r$. The first term under the radical corresponds to the $\alpha^2$ dynamo, while the second term gives rise to the $\alpha\text{--}\Omega$ dynamo. \Citet{Shibata2021} note that when a large-scale dynamo dominates (as we would expect to be associated with dipolar-dominated fields), the $\alpha^2$ term becomes negligible compared to the $\alpha\text{--}\Omega$ term. An unstable mode (with $k_\parallel = k$) will then exist for
\begin{equation}
  k < \kcrit = \parenpow{\bigg}{\frac{\abs{\alpha S_\Omega}}{2 \eta^2}}{1/3}.
\end{equation}

We can estimate this critical wavenumber in our rotating $9\ \msun$ model, using radius $r \sim 20\ \km$, eddy size $L \sim 5\ \km$ (Figure~\ref{fig:3d_vr}), eddy velocity $v \sim 10^3\ \km\ \seconds^{-1}$ (Figure~\ref{fig:velocity}), helicity $h \sim 10^{11}\ \cm\ \seconds^{-2}$ (Figure~\ref{fig:helicity}), and shear $\dd\Omega / \dd r \sim 10^{-3}\ \seconds^{-1}\ \cm^{-1}$. Then $\alpha = \tau h / 3$, with $\tau = L / v$. We approximate magnetic diffusivity to be $\eta \sim \beta = \tau v^2 / 3 \sim 10^{13}\ \cm^2\ \seconds^{-1}$, if we assume turbulent diffusion should be substituted for that due to microscopic conductivity (Figure~\ref{fig:diffusion} in Appendix~\ref{sec:diffusivities}). The result is instability for wavelengths greater than ${\sim}7\ \km$.

As shown in \citet{Shibata2021}, the fastest-growing mode has a wavenumber $\kfast$ that is ${\sim}0.4$ times the critical wavenumber, corresponding to a wavelength of ${\sim}20\ \km$ in our case. The corresponding growth rate is
\begin{equation}
  \Im(\ofast) = \frac{3}{8} \parenpow{\bigg}{\frac{\alpha^2 S_\Omega^2}{2 \eta}}{1/3} \sim 600\ \seconds^{-1}.
\end{equation}
Even considering the approximate nature of these estimates and the simple, linear nature of this analysis,\footnote{The value of $\Im(\ofast)$ likely overestimates the dynamo growth rate during its exponential phase, given that the magnetic field must suffer reconnection in order for its topology to change. This process cannot act faster than the eddy turnover times.} there are plenty of $\ee$-foldings to reach saturation in $1\text{--}2\ \seconds$. In the highly conductive, convective PNS context, even modest rotation will lead to dynamo growth of field; the only question is what the nonlinear saturation strength is and whether rotation is strong enough to produce a dipole-dominated field, both of which we have discussed in \S\ref{sec:dynamo_scaling}.

\begin{figure}
  \centering
  \includegraphics{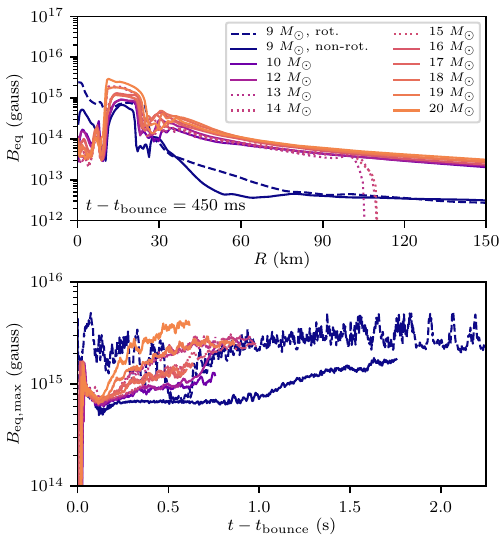}
  \caption{Top:\ profiles of equipartition magnetic field strength obtained by assuming $10\%$ of turbulent kinetic energy is converted into magnetic energy at each radius, calculated $450\ \ms$ after bounce. The dashed line is the $9\ \msun$ rotating model; the dotted lines are models that failed to explode. Bottom:\ peak equipartition field strength inside $150\ \km$ for the same models as a function of time. In both cases, the plotted values are only indicative of potential small-scale field strength. \label{fig:equipartition}}
\end{figure}

We can make a simple estimate of the small-scale magnetic field strengths we expect based on energy considerations. Using only the turbulent velocity field, we can ask what equivalent field has $10\%$ of the fluid's kinetic energy on a given spherical shell at a given time. Radial profiles of this estimate, calculated for all models $450\ \ms$ after bounce, are shown in the upper panel of Figure~\ref{fig:equipartition}. The three models that fail to explode are marked with dotted lines; the $9\ \msun$ rotating model is indicated with a dashed line. All models have a peak equivalent field strength within a factor of a few of $10^{15}\ \gauss$ corresponding to the PNS convection zone at around $15\text{--}20\ \km$, with the rotating model maintaining this strength all the way to the center. In the range $30\text{--}150\ \km$, both of the $9\ \msun$ models have much less turbulent kinetic energy than the other successfully exploding models, while those others show little variation with progenitor mass.

Taking the peak equivalent field strength from each snapshot, we construct time series of maximum magnetic field strengths, shown in the lower panel of Figure~\ref{fig:equipartition}. All our models sustain a region where $10\%$ of the turbulent kinetic energy corresponds to at least $5 \times 10^{14}\ \gauss$ throughout the simulations, which go beyond $1\ \seconds$ after bounce in some cases. Note we are not suggesting that dipole fields would necessarily arise with such magnitudes; estimating the dipole strength requires the Rossby number and convective flux analysis of \S\ref{sec:dynamo_scaling}. Equipartition merely shows that there is sufficient free energy in the system to power magnetar fields, even if much of it only contributes to toroidal fields or to structures that are incoherent on large scales.

% Discussion
\section{Discussion and Conclusions}
\label{sec:discussion}

The observed NS population displays a wide range of magnetic field strengths and spins. While to some degree evolution after the birth of the NS can lead to different outcomes (most notably the millisecond pulsars that have undergone accretion from a companion as described in \citet{Bhattacharya1991} or \citet{Phinney1994}), much of the diversity must be attributed to the range of conditions before and during core collapse. In particular, the dichotomy between magnetars and the standard radio pulsars demands explanation.

There are a number of mechanisms that can amplify the magnetic fields in the cores of massive stars during collapse, resulting in pulsar or possibly magnetar strengths. Flux freezing, the MRI, and dynamos have all been considered, and it is likely that each plays some role in Nature. Flux freezing has difficulty attaining magnetar fields, however, and both it and the MRI do not naturally result in a dichotomy of outcomes given a monomodal distribution of initial conditions.

On the other hand, convective dynamos can produce fields with a natural dichotomy of outcomes. As first realized in the context of planetary bodies, systems with modified Rossby numbers $\rol$ below a threshold of ${\sim}0.12$ acquire predominantly dipolar fields, while systems with relatively slower rotation (or faster convective motions) become multipolar, with relatively weaker dipoles \citep{Christensen2006,Olson2006}. The behavior near the transition is less clear;\footnote{Near the transition, simulations predict dipoles undergoing occasional reversals \citep{Christensen2006}. Observations of M~dwarfs show evidence for bistability of dipolar and multipolar dynamos \citep{Morin2011}. \Citet{Brun2015} note this might only happen for low stratification and large magnetic Prandtl numbers, the former of which does not apply to PNS convection.} this does not qualitatively affect our arguments, though should be considered in future work that quantifies the distribution of PNS Rossby numbers.

In the low-Rossby-number regime, the resulting dipole strength has a simple scaling with convective flux that is independent of the classic dimensionless numbers, and this has been extended to and tested against T~Tauri stars and old M~dwarfs \citep{Christensen2009}. The Rossby-number dichotomy motivates us to analyze core-collapse models in the same framework, applying the \citet{Christensen2009} scalings to a new physical regime, that of a PNS at the center of a core-collapse supernova.

It has already been shown that vigorous PNS convection is generic in core collapse \citep{Dessart2006,Nagakura2020}, and so there will always be interaction between this motion and the magnetic fields present in the progenitor. Moreover, even a moderately rotating progenitor, such as our $9\ \msun$ model, can attain $\rol < 0.12$ (Figure~\ref{fig:rossby}) in and around the PNS. That is, the extreme rotation at an appreciable fraction of breakup sometimes employed to maximize magnetic growth may be unnecessary for producing large fields.

We note \citet{Masada2021} find that even very slow rotation (periods longer than $160\ \ms$) can produce large dipolar fields rapidly (${\sim}6\text{--}10\ \ms$ per $\ee$-folding) given large eddies (and thus long turnover times) in deep-core convection. All of their simulations show larger convective cells than we see, and it is unclear how generic this behavior is, or whether it might be the result of, e.g., the way in which they impose $\ye$ profiles to capture neutrino effects. Their results illustrate how important the details of PNS convection can be for generating NS magnetic fields.

A somewhat different perspective is given by \citet{Raynaud2020}, whose simulations differ from those of \citet{Masada2021} by (1) being anelastic; (2) using heat flux rather than composition to drive convection; (3) using generally larger momentum, thermal, and magnetic diffusivities; and (4) employing very rapid rotation (with periods approaching breakup at $2\ \ms$). \Citet{Raynaud2020} find magnetic energy densities exceeding kinetic energy densities, with saturated field strengths scaling with rotation rate, from which they infer magnetostrophic rather than turbulent force balance dominates \citep[see][]{Brun2015}. Their fields grow at a rate of ${\sim}70\ \ms$ per $\ee$-folding. Rapid rotation might be sufficient to generate magnetar fields without the aid of convective flux, but it need not be necessary. Indeed, requiring all magnetars to be rotating near breakup at birth would lead to observable effects in supernovae at very early \citep{Burrows2007} or slightly later \citep{Kasen2010} times, which would be in tension with the low observed rates of particularly luminous transients.

Our own suite of radiation-hydrodynamic simulations, though lacking magnetic fields, employ full neutrino transport and a nuclear equation of state, and therefore should develop particularly realistic convection. With progenitors ranging from $9\ \msun$ to $25\ \msun$, they show sufficient convective flux (Figure~\ref{fig:convection_3d}) to support even magnetar fields according to the \citet{Christensen2009} scalings (Figure~\ref{fig:expected_dipole}). Predicted field strength increases slightly with progenitor mass, consistent with more massive progenitors leaving behind a more massive PNS, which subsequently has more convective flux \citep{Nagakura2020}.

That there is enough kinetic energy in PNS convection to correspond to strong magnetic fields in equipartition (Figure~\ref{fig:equipartition}) is not new \citep{Endeve2010,Endeve2012,Obergaulinger2014}. This fact suggests that on small scales magnetar-magnitude fields are naturally generated in every NS at birth. However, the dynamo studies of \citet{Christensen2006}, \citet{Olson2006}, and \citet{Christensen2009} provide reason to believe that when sufficient rotation is present, large-scale, dipolar fields of magnetar magnitude will in fact arise. This connection is strengthened by the presence of large values of helicity in the regions of interest (Figures~\ref{fig:3d_helicity} and~\ref{fig:helicity}), which drives large-scale field growth in mean-field theory \citep{Charbonneau2014}.

The scaling results we employ only predict the state of the magnetic field once it has reached nonlinear saturation. In order for convective dynamos to be relevant to NS field amplification, there must be sufficient time for field growth while convection lasts (i.e., before the core has fully deleptonized and cooled). Rough estimates of relevant timescales (see \S\ref{sec:dynamo_scaling}), as well as more detailed linear mode analysis (\S\ref{sec:convection}), indicate that there is ample time for exponential growth of at least a factor of $10^6$ in field strength. The $9\ \msun$ models we analyze are run to ${\sim}2\ \seconds$ after bounce, and, while convection may begin to diminish by this time (Figure~\ref{fig:luminosities}), conditions are still amenable to the generation of strong dipolar fields in the low-Rossby-number regime.

Our focus has been on using dynamo scaling relations to inform PNS field generation. We note, however, that data for PNS fields can inform the general study of dynamos. The objects we consider have very different diffusion physics compared to planetary interiors and low-mass stars, with particularly high conductivities and with neutrino-mediated momentum and thermal diffusivities. While some dimensionless characterizations of our models (e.g.\ Rossby, Reynolds, and Prandtl numbers) are roughly commensurate with studies in different regimes, others (e.g.\ magnetic Reynolds and magnetic Prandtl numbers $\rem$ and $\prm$) are much larger. We refer the reader to Appendix~\ref{sec:diffusivities} for more details about the diffusivities and dimensionless numbers in our models.

These differences in dimensionless numbers may be relevant according to \citet{Brandenburg2014}, who finds that the ratio of kinetic to magnetic dissipation in simulations is roughly $\prm^{1/3}$. Extrapolating this result from the largest value in that study, $\prm = 20$, to $\prm \sim 10^{14}$ applicable to PNS convection, we might expect $\fohm \sim 10^{-5}$ in Equation~\eqref{eq:dipole}, meaning the expected dipole might be reduced by a factor of ${\sim}10^2$ from what is plotted in Figure~\ref{fig:expected_dipole}. We emphasize, however, that this extrapolation is uncertain, and in fact \citet{Brandenburg2014} mentions that it may be that at large $\rem$ (which is much larger in a PNS than in any simulation), the dissipation ratio may lose its dependence on $\prm$. Moreover, as noted by \citet{Thompson1993}, the nominal neutrino viscous damping scale is comparable to the neutrino mean free path, so any turbulent cascade can continue down to much smaller scales set by electron viscosity, where the magnetic Prandtl number is much closer to unity.

Though explicit numerical tests with realistic magnetic diffusivities are intractable, we assume, and there is reason to believe \citep{Christensen2009}, that for sufficiently large dimensionless numbers the behavior of dynamos should enter an asymptotic regime and lose its dependence on these numbers. This assumption, and indeed the range of validity for dynamo scaling, will be tested over time as more PNS simulations are compared to NS observations.

In conclusion, we find it plausible that PNS convection plays a large role in determining salient properties of NSs at birth. Convective motions can further augment fields that have already been amplified via flux freezing. If sufficient rotation exists---and we emphasize that ``sufficient'' need not be close to breakup---dynamo action will lead to a dipolar field with a strength scaling with convective flux. This channel can explain the magnetars, as the strong dipole field will naturally remove angular momentum on short timescales, leaving a slowly rotating, highly magnetized NS to be observed. Without sufficient rotation, the convective regions still exist, but no magnetar-like dipole will be created, and the resulting NS will take longer to spin down, explaining standard radio pulsars.

There is important work to be done to be sure this model holds together. While we can be sure that strong convection is common in proto-neutron stars, it is still unclear what distribution of rotational profiles and seed magnetic fields we should expect in the late stages of massive star evolution. In addition, future PNS modeling would benefit from a more comprehensive treatment of the physics. The models we study here excel at hydrodynamics and neutrino transport, including a sophisticated equation of state and detailed neutrino-matter interactions, but they do not include the (generally energetically subdominant) magnetic fields we know must be present. The most recent PNS/magnetic-field studies by various groups \citep[e.g.,][]{Franceschetti2020,ReboulSalze2021,Raynaud2020,Masada2021} have begun to address magnetic field generation through dynamos or the MRI, though at the expense of approximating the physics that drives PNS convection. What is needed in the next theoretical phase is the merger of state-of-the-art 3D supernova codes incorporating detailed neutrino transfer, equations of state, and initial models with a sophisticated treatment of magnetic fields. The resulting radiation-magnetohydrodynamic code will enable evolutionary simulations of the 3D PNS cores birthed by supernovae and carried to late times. The ultimate goal is to determine the mapping between the initial massive-star progenitors and the final masses, spins, magnetic dipoles, and magnetic multipolarity structures of the residual NSs. Such a theory should usefully connect the radio-pulsar and magnetar communities with the supernova theory community studying the creation of the NSs upon which the former focus. This is a long-awaited goal that we suggest may soon be within reach.

% Acknowledgments
\acknowledgments

\strut

We thank Eric Blackman, Romain Teyssier, Tianshu Wang, and Hiroki Nagakura for fruitful discussions and Joe Insley of Argonne National Laboratory for the graphical rendering of Figure~\ref{fig:3d_pns}. We acknowledge support from the U.~S.\ Department of Energy Office of Science and the Office of Advanced Scientific Computing Research via the Scientific Discovery through Advanced Computing (SciDAC4) program and Grant DE-SC0018297 (subaward 00009650) and support from the U.~S.\ National Science Foundation (NSF) under Grants AST-1714267 and PHY-1804048 (the latter via the Max-Planck/Princeton Center (MPPC) for Plasma Physics). The bulk of the computations presented in this work were performed on Blue Waters under the sustained-petascale computing project, which was supported by the National Science Foundation (awards OCI-0725070 and ACI-1238993) and the state of Illinois. Blue Waters was a joint effort of the University of Illinois at Urbana--Champaign and its National Center for Supercomputing Applications. The two $9\ \msun$ models were simulated on the Frontera cluster (under awards AST20020 and AST21003), and this research is part of the Frontera computing project at the Texas Advanced Computing Center \citep{Stanzione2020}. Frontera is made possible by NSF award OAC-1818253. Additionally, a generous award of computer time was provided by the INCITE program, enabling this research to use resources of the Argonne Leadership Computing Facility, a DOE Office of Science User Facility supported under Contract DE-AC02-06CH11357. Finally, the authors acknowledge computational resources provided by the high-performance computer center at Princeton University, which is jointly supported by the Princeton Institute for Computational Science and Engineering (PICSciE) and the Princeton University Office of Information Technology, and our continuing allocation at the National Energy Research Scientific Computing Center (NERSC), which is supported by the Office of Science of the U.~S.\ Department of Energy under contract DE-AC03-76SF00098.

% Software
\software{\code{Fornax} \citep{Skinner2019,Vartanyan2019}, \code{VisIt}}

% References
\bibliographystyle{aasjournal}
\bibliography{references}

% Diffusivities and Dimensionless Numbers
\appendix
\section{Diffusivities and Dimensionless Numbers}
\label{sec:diffusivities}

In a turbulent, convective, conducting fluid, we generally expect the character of the flow to depend on three diffusivities:\ those of momentum, $\nu$ (i.e.\ the kinematic viscosity); thermal energy, $\kappa$; and magnetic field, $\eta$. Viscosity determines how far down in scale turbulence will be driven, and, in combination with thermal diffusivity, determines in many contexts whether convection drives turbulence. Some magnetic diffusivity is required for successful dynamos, allowing reconnection to change the field topology. Too much, however, prevents magnetic fields from being sustained.

\Citet{Burrows1988} note that $\nu$ \citep[see][]{VanWeert1983} and $\kappa$ \citep[see][]{VandenHorn1984} are dominated in the optically thick regions by the effects of neutrinos diffusing through the fluid. \Citet{Thompson1993} approximate the scalings
\begin{equation}
  \nu_\mathrm{diff} = 2 \times 10^9\ \rho_{14}^{-4/3} T_{30} f(\ye)^{-1}\ \cm^2\ \seconds^{-1}
\end{equation}
and
\begin{equation}
  \kappa_\mathrm{diff} = 1.5 \times 10^{10}\ \rho_{14}^{-2/3} T_{30}^{-1}\ \cm^2\ \seconds^{-1},
\end{equation}
where $\rho_{14} = \rho / (10^{14}\ \g\ \cm^{-3})$, $T_{30} = \kb T / (30\ \mev)$, and $f(\ye) = (\ye^{1/3} + (1 - \ye)^{1/3})^{-1}$. Outside the neutrinospheres, however, the neutrino mean free paths begin to grow larger than the length scales of interest, and so the effective diffusivities must be modified. As is done in \citet{Guilet2015} in the context of MRI growth, for the purposes of our estimations we define
\begin{equation}
  \nu = \frac{\nu_\mathrm{diff}}{1 + (\lambda / H_p)^2}
\end{equation}
and
\begin{equation}
  \kappa = \frac{\kappa_\mathrm{diff}}{1 + (\lambda / H_p)^2},
\end{equation}
where
\begin{equation}
  \lambda = 200\ \rho_{14}^{-1/3} T_{30}^{-3} f(\ye)\ \cm
\end{equation}
\citep{Thompson1993} and $H_p = -p / (\dd p / \dd r)$ is the pressure scale height at a given radius.

\begin{figure}
  \centering
  \includegraphics{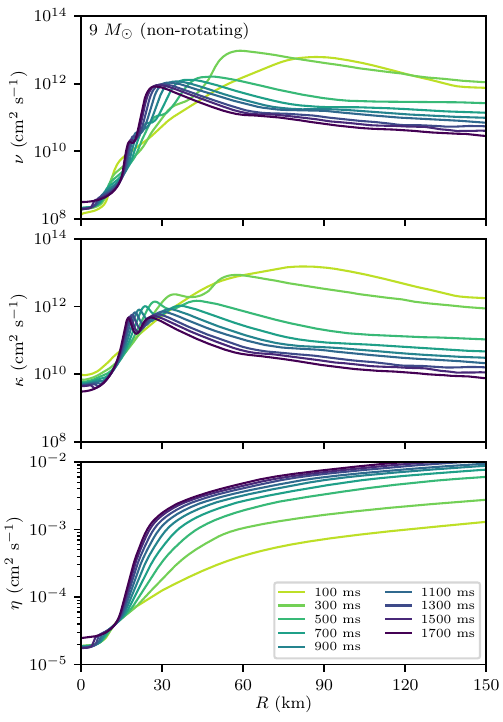}
  \caption{Profiles of momentum diffusivity (kinematic viscosity), thermal diffusivity, and magnetic diffusivity for the non-rotating $9\ \msun$ model. The first two are corrected for long neutrino mean free paths, dividing the optically thick value by $1 + (\lambda / H_p)^2$, where $\lambda$ is the neutrino mean free path and $H_p$ is the pressure scale height. Colors correspond to time after bounce. \label{fig:diffusion}}
\end{figure}

Angle-averaged radial profiles of these diffusivities, calculated from our $9\ \msun$ model at various times, are shown in Figure~\ref{fig:diffusion}. In the core of the PNS (approximately the inner $30\ \km$), $\nu$ ranges from $10^8$ to $10^{12}\ \cm^2\ \seconds^{-1}$, while $\kappa$ ranges from $10^{10}$ to $10^{12}\ \cm^2\ \seconds^{-1}$. They both range from $10^{11}$ to $10^{13}\ \cm^2\ \seconds^{-1}$ in the region exterior to this (inside $150\ \km$).

Magnetic diffusivity arises from electrical resistivity, which is taken to be dominated by electron-proton scattering. Following \citet{Thompson1993}, we define
\begin{equation}
  \eta = \frac{1}{3 \times 10^{4}}\ \rho_{14}^{-1/3} (5 \ye)^{-1/3}\ \cm^2\ \seconds^{-1}.
\end{equation}
Radial profiles of $\eta$ are also shown in Figure~\ref{fig:diffusion}, where it varies from $10^{-5}$ to $10^{-3}\ \cm^2\ \seconds^{-1}$ inside the PNS, increasing up to $10^{-2}$ in the exterior region. If we let $\vturb$ be the RMS velocity of turbulence, subtracting off bulk radial motion at each radius, and, in the case of rotating models, net rotation at each radius and latitude, and if we define $\vave^r = \ave{\rho v}_\Omega / \ave{\rho}_\Omega$, $\vave^\theta = 0$, and $\vave^\phi = 0$ (non-rotating models) or $\vave^\phi = \ave{\rho v}_\phi / \ave{\rho}_\phi$, then
\begin{equation}
  \vturb = \ave[\big]{(\vprime)^2}_\Omega^{1/2},
\end{equation}
where $\vprime = \vec{v} - \vavevec$. This provides a characteristic velocity scale.

Next, we define a characteristic turbulent length scale as follows. Decompose each component of the turbulent velocity field via real spherical harmonics $Y_{\ell m}$, calculating the power in each harmonic degree $\ell$ up to a practical maximum of $\lmax = 32$, and finding the average of $\ell$ weighted by this power:
\begin{equation}
  \vec{v}_{\ell m} = \oint (\vec{v} - \vavevec) Y_{\ell m} \, \dd\Omega,
\end{equation}
with
\begin{equation}
  \vec{v}_\ell = \sum_{m = -\ell}^\ell \vec{v}_{\ell m} Y_{\ell m}
\end{equation}
and
\begin{equation}
  \bar{\ell} = \frac{\sum_{\ell = 1}^{\lmax} \ell \ave{v_\ell^2}_\Omega}{\sum_{\ell = 1}^{\lmax} \ave{v_\ell^2}_\Omega}.
\end{equation}
Take the length scale to be
\begin{equation}
  \lturb = \frac{\pi R}{\bar{\ell}},
\end{equation}
where $R$ is the spherical radius.

\begin{figure}
  \centering
  \includegraphics{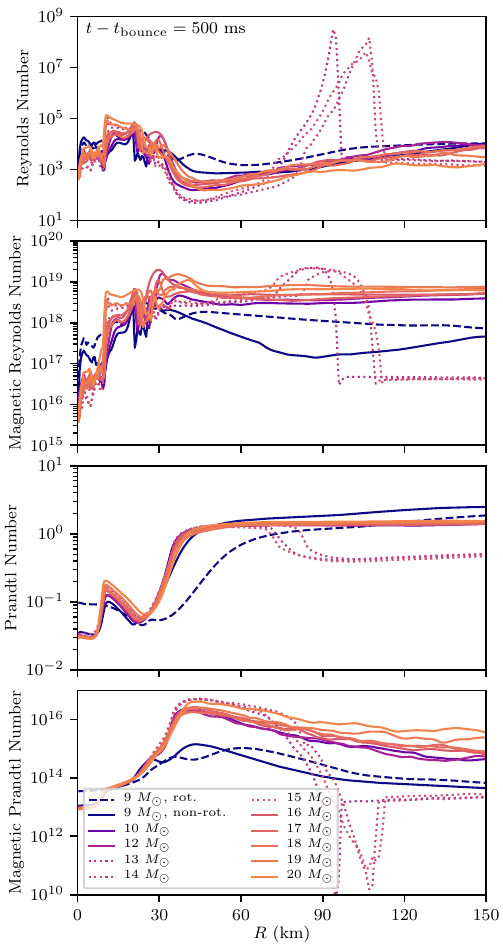}
  \caption{Profiles of dimensionless numbers for various models $500\ \ms$ after bounce, defined as the Reynolds number $\re = \vturb \lturb / \nu$, magnetic Reynolds number $\rem = \vturb \lturb / \eta$, Prandtl number $\pr = \nu / \kappa$, and magnetic Prandtl number $\prm = \nu / \eta$. Here, $v$ is the RMS turbulent velocity, $L$ is the characteristic turbulent length scale, and $\nu$ and $\kappa$ are adjusted for neutrino mean free paths as in Figure~\ref{fig:diffusion}. The dashed line is the rotating model; the dotted lines are models that failed to explode, with sharp features occurring at the stalled shock. \label{fig:dimensionless}}
\end{figure}

Dimensionless numbers constructed from diffusivities provide important characterizations of the fluid dynamics. The Reynolds number
\begin{equation}
  \re = \frac{\vturb \lturb}{\nu},
\end{equation}
averaged over angle in a given snapshot, is plotted in the first panel of Figure~\ref{fig:dimensionless} for the suite of models. It lies between $10^3$ and $10^5$ in all models within the PNS core, delineated by a sharp change in density, and is perhaps an order of magnitude lower than this immediately exterior to the core. Curiously, $\re$ is not so large as to prohibit resolution of the viscous scale in future simulations. The three models that do not explode (those with $13$, $14$, and $15\ \msun$ progenitors \citep{Burrows2020}), have clearly distinct $\re$ profiles (and indeed profiles of other dimensionless numbers). Otherwise, there is little variation of $\re$ with progenitor mass.

The magnetic Reynolds number can be defined similarly:
\begin{equation}
  \rem = \frac{\vturb \lturb}{\eta}.
\end{equation}
As shown in Figure~\ref{fig:dimensionless}, $\rem$ is always above $10^{16}$, indicating a highly conducting plasma where magnetic diffusion at characteristic turbulent scales is quite small. Excluding models that do not explode, only the $9\ \msun$ models are outliers in $\rem$, with the rotating model having larger values interior to $30\ \km$ and the non-rotating model having lower values exterior to this. We do not, however, expect the differences to result in qualitative differences in terms of turbulence and dynamo generation, given how large $\rem$ is in all cases.

We can also construct the Prandtl number
\begin{equation}
  \pr = \frac{\nu}{\kappa}
\end{equation}
and magnetic Prandtl number
\begin{equation}
  \prm = \frac{\nu}{\eta},
\end{equation}
shown in the bottom panels of Figure~\ref{fig:dimensionless}. While the former is close to unity over the region of interest, the large conductivity of the fluid results in values of $\prm$ ranging from $10^{13}$ to $10^{16}$.

\end{document}